\documentstyle[prl,aps,epsfig,twocolumn]{revtex}

\begin{document}

\draft

\twocolumn[\hsize\textwidth\columnwidth\hsize\csname@twocolumnfalse\endcsname
\title{Fast entanglement of two charge-phase qubits \\
through nonadiabatic coupling to a large junction}
\author{Y. D. Wang, P. Zhang, D. L. Zhou, C. P. Sun$^{a,b}$}
\address{Institute of Theoretical Physics, The Chinese Academy of Sciences,
Beijing, 100080, China}
\date{\today}
\maketitle
\begin{abstract}
We propose a theoretical protocol for quantum logic gates between
two Josephson junction charge-phase qubits through the control of
their coupling to a large junction. In the low excitation limit of
the large junction when $E_{J}\gg E_{c}$, it behaves effectively
as a quantum data-bus mode of a harmonic oscillator. Our protocol
is efficient and fast. In addition, it does not require the
data-bus to stay adiabatically in its ground state, as such it can
be implemented over a wide parameter regime independent of the
data-bus quantum state.\\

\end{abstract}
 ]

Significant progress has been made in improving the quantum coherence of
Josephson junction (JJ) based qubits \cite{Han1,vion1} since the first
experiment breakthrough in 1999 \cite{ch}. The coherent interaction of two
JJ qubits has been implemented and the initial indications of their
entanglement have been detected \cite{twoqs,lobb}. Even the demonstration of
2-bit conditional gate of charge qubits was reported most recently \cite%
{cnot}. These developments have paved the way towards the realization of the
two important elements of universal quantum computation; the ability to
implement arbitrary single-bit rotations and the controlled logic gates
between two qubits. Various theoretical schemes have been proposed for
quantum logic gate operations of JJ qubits. The most important measure of
their success depends on the proposed mechanism for a controlled coupling
between the JJ qubits. Most protocols involve the coupling of each JJ qubit
to an auxiliary data-bus (such as variable transformer) and affect an
effective coupling between two qubits through the elimination of dynamic
variables of the data-bus \cite{maklin1,averin1,Han2,kim,zagoskin,you,you2}.

The existing schemes
\cite{maklin1,averin1,Han2,kim,zagoskin,you,you2} usually apply to
weak coupling parameters, thus leading to slow gates with long
operation times. Most require the data-bus to be adiabatic, i.e.
to stay in its ground state or a different pure state during gate
operation. In this article, we suggest a protocol capable of
efficient and fast gate operations between two charge-phase qubits
\cite{vion1}. As will be shown  in detail below, our protocol is
insensitive to the state of the data-bus, thus is not restricted
to the weak coupling limit as required by the adiabatic condition.
In fact, the resulting strong effective coupling leads to short
gate operation times. Furthermore, all control parameters in our
protocol can be modulated within current experiments.

Our main idea is to construct a system of two JJ qubits linked by a quantum
data-bus of a harmonic oscillator with the creation (annihilation) operator $%
\hat{a}\left( \hat{a}^{\dag }\right) $. Each of the two qubits is linearly
coupled to this data-bus mode and described by the following form of the
Hamiltonian $H\left( t\right) =\hat{a}\hat{F}(t)+h.c.$ with a general force $%
\hat{F}(t)=$ $\beta _{1}(t)\hat{F}_{1}+\beta _{2}(t)\hat{F}_{2}$. $\hat{F}%
_{j}\ (j=1,2)$ is a dynamic variable (not necessarily an observable) of the
j-th qubit. It is known that when $\hat{F}_{j}$ commutes with its own
conjugate $\hat{F}_{j}^{\dag }$, the time evolution of $H(t)$ is given by
\begin{equation}
U(t)=\exp \left[ i{\sum\limits_{k,j}}\mu _{k,j}(t)\hat{F}_{k}^{\dag }\hat{F}%
_{j}\right] e^{\alpha (t)a}e^{\alpha ^{\ast }(t)a^{\dag }},
\end{equation}%
where the expressions for $\mu _{kj}(t)$ and $\alpha (t)$ can be explicitly
obtained using the Wei-Norman algebra \cite{weinorman,sun1}. The crucial
observation is, at certain instants of time $t=T$, $\alpha (T)=0$ and thus
the time evolution is simply described by
\begin{equation}
U(T)=\exp \left[ i{\sum\limits_{k,j}}\mu _{kj}(T)\hat{F}_{k}^{\dag }\hat{F}%
_{j}\right] .
\end{equation}%
Formally, the quadratic terms of $F_{j}$ correspond to the effective
nonlinear interactions between the two qubits. It is important to note that
even for a fixed time $T$, there still exists enough parameters in the
system to modulate $\mu _{kj}(T)$ to affect the required logic gate. Clearly
our strategy does not require the state of the data-bus to be adiabatic.
Similar approaches have been adopted before in protocols of
\textquotedblleft quantum computing with thermal ions" \cite{Molmer} and by
Wang {\it et. al.} \cite{wang}.

\begin{figure}[ptb]
\begin{center}
\includegraphics[width=6cm,height=4cm]{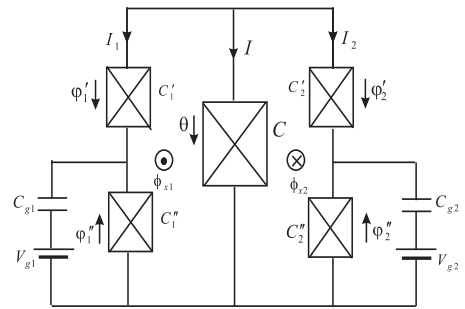}
\end{center}
\caption{Schematic diagram of Two charge-phase qubits are coupled through a
large junction. }
\end{figure}

We consider a system of coupled charge-phase qubits as illustrated by the
electronic circuit shown in Fig. 1. Similar systems have been recently
discussed by You {\it et. al.} \cite{you}. Our idea works in the limit when
the large junction (of capacitance $C$) stays at a low excitation or even a
thermal state. The same model with only one qubit was used to demonstrate
the progressive decoherence by us \cite{sun2}. Each of the Cooper pair box
is split into two small junctions of capacitance $C_{k}^{^{\prime }}$ and $%
C_{k}^{^{\prime \prime }}$ ($k=1,2$) that form a superconducting loop. $%
C_{gk}$ is the capacitance of the gate, and $E_{Jk}$ is the Josephson
coupling energy. For simplicity we set $C_{k}^{^{\prime }}=C_{k}^{^{^{\prime
\prime }}}\equiv C_{k}$ and $E_{Jk}^{^{\prime }}=E_{Jk}^{^{\prime \prime
}}\equiv E_{Jk}$. If $C$ is much larger than all other capacitances in the
circuit, then the Coulomb energy can be approximated by a non-entangled
form. The total Hamiltonian containing the reduced Coulomb energy and the
three Josephson coupling energies, then reads
\begin{eqnarray}
H &=&\sum\limits_{k=1,2}\left[ E_{ck}(n_{k}-n_{gk})^{2}-E_{Jk}\left( \cos
\varphi _{k}^{^{\prime }}+\cos \varphi _{k}^{^{^{\prime \prime }}}\right) %
\right]  \nonumber \\
&&+E_{c}N^{2}-E_{J}\cos \theta ,  \label{hs1}
\end{eqnarray}%
where $E_{ck}={2e^{2}}/{C_{\Sigma k}}$, $n_{gk}=C_{gk}V_{gk}/{2e}$, $E_{c}={%
2e^{2}}/{C}$, and $C_{\Sigma k}=C_{gk}+C_{k}^{^{\prime }}+C_{k}^{^{\prime
\prime }}$. Here $n_{k}$ is the number of Cooper pairs on the k-th island,
while $N$ is the number of Cooper pairs on the Coulomb island connected with
the large junction. $\varphi _{k}^{^{\prime }}$, $\varphi _{k}^{^{\prime
\prime }}$, and $\theta $ superconduction phase differences across the
relevant junctions. They are related through the fluxoid quantization
condition around the loop $\theta +\varphi _{k}^{^{\prime \prime }}-\varphi
_{k}^{^{\prime }}=2\Theta _{k}$ and $\Theta _{k}=\pi {\Phi _{xk}}/{\phi _{0}}
$. Introducing $\varphi _{k}=(\varphi _{k}^{^{\prime }}+\varphi
_{k}^{^{\prime \prime }})/{2}$, we rewrite Eq. (\ref{hs1}) to
\begin{eqnarray}
H &=&\sum\limits_{k=1,2}E_{ck}(n_{k}-n_{gk})^{2}+E_{c}N^{2}-E_{J}\cos \theta
\nonumber \\
&&-2\sum\limits_{k=1,2}E_{Jk}\cos (\frac{\theta }{2}-\Theta _{k})\cos
\varphi _{k},
\end{eqnarray}%
now with the quantization condition $\left[ \varphi _{k},n_{l}\right]
=i\delta _{lk}$. As is well known for charge-phase qubit, when $n_{gk}=0.5$
and when $E_{jk}$ is not much larger than $E_{ck}$, the linear combinations
of the two lowest charge eigenstates $\{|0\rangle _{i}$,$|1\rangle _{i}\}$
for each of the split Cooper pair box consist a good representation of a
qubit.

We now consider the\textquotedblleft coherent" regime when
$E_{J}\gg E_{c}$. As was found before in the study of quantum
phase transitions \cite{zurek} and also some other references such
as \cite{you}, the spectrum of the low energy part of the large
junction can be described approximately by a harmonic oscillator.
Within this approximation, we expand the above Hamiltonian around
$\theta =0$ up to $O(\theta ^{2})$ and obtain an effective
spin-boson Hamiltonian
\begin{eqnarray}
H &=&H\left( g_{k},f_{k}\right)   \nonumber  \label{3} \\
&\equiv &\sum\limits_{k=1,2}\left[ g_{k}(a^{\dag }+a)\sigma
_{xk}+f_{k}\sigma _{xk}\right] +\Omega a^{\dag }a,
\end{eqnarray}%
where
\begin{eqnarray}
\Omega  &=&\sqrt{2E_{c}E_{J}},  \nonumber \\
g_{k} &=&-\left( \frac{E_{c}}{32E_{J}}\right) ^{\frac{1}{4}}E_{Jk}\sin
\Theta _{k},  \nonumber \\
f_{k} &=&-E_{Jk}\cos \Theta _{k},
\end{eqnarray}%
and the quasi-spin and boson operators are defined as
\begin{eqnarray}
\sigma _{xk} &=&|1\rangle _{kk}\langle 0|+|0\rangle _{kk}\langle 1|,
\nonumber \\
a &=&\left( \frac{E_{J}}{8E_{c}}\right) ^{\frac{1}{4}}\theta +i\left( \frac{%
E_{c}}{2E_{J}}\right) ^{\frac{1}{4}}N.
\end{eqnarray}%
The validity of this approximation can be verified through a
straightforward  numerical calculation. In fact, we first
calculate the evolution wave function $\left\vert \psi
_{1}\right\rangle $\ , which is  governed by the approximate
Hamiltonian (5). Then we compare it with the exact wave function
$\left\vert \psi _{2}\right\rangle $\  governed by the exact
Hamiltonian (3) to a sufficiently high precision  in expansion of
the cosine function of $\theta $\ (eg, up to $ O(\theta ^{6})$).
In this way  we showed  that,  during one gate operation time (
$\sim 10^{-10} $s as calculated in this paper later), the
deviation of $\left\vert \psi
_{1}\right\rangle $\ from $\left\vert \psi _{2}\right\rangle $\ is $%
10^{-4}$ at most.

This equivalent spin-boson system can be solved exactly since the
interaction term $g_{k}(a^{\dag }+a)\sigma _{xk}$ commutes with the free
spin-part. For each spin eigenstate, the $\sigma _{xk}$ acts as a linear
external force on the boson part. With the Wei-Norman algebraic method \cite%
{weinorman}, we find in the interaction picture%
\begin{eqnarray}
U_{I}(t) &\equiv &U_{I}[t;g_{k}]=\exp [-iC\left( t\right) -iA(t)\sigma
_{x1}\sigma _{x2}]  \nonumber \\
&&\times \prod\limits_{k=1,2}\exp [-iB_{k}(t)a\sigma _{xk}]\exp
[-iB_{k}^{\ast }(t)a^{\dag }\sigma _{xk}],  \label{kk}
\end{eqnarray}%
where the time-dependent parameters are given by
\begin{eqnarray}
B_{k}(t) &=&\frac{g_{k}}{-i\Omega }\left( e^{-i\Omega t}-1\right) ,
\nonumber \\
A\left( t\right) &=&\frac{2g_{1}g_{2}}{\Omega }\left( \frac{1}{i\Omega }%
\left( e^{i\Omega t}-1\right) -t\right) ,  \nonumber \\
C\left( t\right) &=&\frac{g_{1}^{2}+g_{2}^{2}}{\Omega }\left( \frac{1}{%
i\Omega }\left( e^{i\Omega t}-1\right) -t\right) .
\end{eqnarray}%
When $B_{k}(t)=0$, the evolution operator $U_{I}\left( t\right) \equiv \exp
[-iC\left( t\right) -iA(t)\sigma _{x1}\sigma _{x2}]$ becomes independent of
variables of the large junction. It takes the canonical form capable of
two-qubit gate operations. We note that $B_{i}(t)$ is a periodic function of
time and it vanishes at $t_{n}={2n\pi }/{\Omega }$ for integer $n=0,\pm
1,\pm 2,\cdots $. At these instants of time, the time evolution operator
becomes explicitly as
\[
U_{I}\left( t_{n}\right) =\exp \left[ -i\frac{4n\pi g_{1}g_{2}}{\Omega ^{2}}%
\sigma _{x1}\sigma _{x2}\right] ,
\]%
up to a phase factor $\exp \left[ -i2n\pi {(g_{1}^{2}+g_{2}^{2})}/{\Omega
^{2}}\right] $ in the interaction picture. This is equivalent to a system of
two coupled qubits with an interaction of the form $\propto \sigma
_{x1}\sigma _{x2}$.

For a two-qubit system governed by a Hamiltonian of the form $g\sigma
_{x1}\sigma _{x2}$, one usually adjusts the evolution time to realize an
arbitrary two-bit controlled rotation $U_{xx}\left( \xi \right) \equiv \exp
[-i\xi \sigma _{x1}\sigma _{x2}]$ at $t={\xi }/{g}$. In our system, the
evolution time is fixed at $t_{n}$ by the requirement of $B_{i}(t_{n})=0$.
This is in fact not a problem as experimentally one can vary $g_{1}$ and $%
g_{2}$ to affect the desired rotation $U_{xx}(\xi )$ at $t_{n}$. We note
that $g_{i}$ depends on the external flux $\Phi _{xk}$ of the loop $\Theta
_{k}=\pi {\Phi _{xk}}/{\phi _{0}}$, so it is easy to adjust to the maximum
of $g_{k\max }=g_{k}\left( \Theta _{k}=\frac{\pi }{2}\right) =-\left( \frac{%
E_{c}}{32E_{J}}\right) ^{\frac{1}{4}}E_{Jk}$. The minimal time for one
operation $t_{n\min }=\frac{2n_{\min }\pi }{\Omega }$ is given by
\begin{equation}
n_{\min }=\left[ \frac{\Omega ^{2}}{2g_{1}\left( \frac{\pi }{2}\right)
g_{2}\left( \frac{\pi }{2}\right) }\right] +1,
\end{equation}%
where $[x]$ denotes the integer part of $x$.

At the same time, we note that the free Hamiltonian part $H_{0}=\sum
f_{k}\sigma _{xk}$ does not vanish during the above discussed two-qubit
operation. However, it commutes with the interaction term $\propto \sigma
_{x1}\sigma _{x2}$ and simply leads the evolution operator being
\begin{equation}
U_{xx}\left( \xi \right) =[e^{-i\xi \sigma _{x1}\sigma _{x2}}e^{-i\Sigma
f_{k}\left( \xi \right) \sigma _{xk}}]e^{i\Sigma f_{k}\left( \xi \right)
\sigma _{xk}},
\end{equation}%
i.e. augmented by two single-bit operations with $f_{k}\left( \xi \right)
=-E_{Jk}\cos \Theta _{k}\left( \xi \right) $. $\Theta _{k}\left( \xi \right)
$\ satisfies the equation
\begin{equation}
\frac{n\pi }{4\Omega ^{2}}\left( \frac{8E_{c}}{E_{J}}\right) ^{\frac{1}{2}%
}E_{J1}E_{J2}\sin \Theta _{1}\sin \Theta _{2}=\xi .
\end{equation}

Alternatively we can remove the influence of the large junction on the
two-qubit logic operation by using two operations in succession, a concept
as presented recently for a two-qubit gate on trapped ions \cite{zheng}. The
two steps are described as follows.

First, we evolve the system with the Hamiltonian $H$ for a duration ${\tau }/%
{2}$. The evolution operator (in the interaction picture) is
\begin{equation}
U_{I}\left( \frac{\tau }{2}\right) =U_{I}\left[ \frac{\tau }{2};g_{k}\right]
.
\end{equation}

Second, at time $t={\tau }/{2}$, we instantly reverse the direction of the
magnetic field such that a sudden change of flux from $\Phi _{xk}$ to $-\Phi
_{xk}$ occurs that leads to $g_{k}^{^{\prime }}=-g_{k}$ and $f_{k}^{^{\prime
}}=f_{k}$. The system is then driven by a new Hamiltonian $H^{\prime
}=H\left( -g_{k},f_{k}\right) $, with which we evolve for another ${\tau }/{2%
}$. Since $H_{0}=H_{0}^{^{\prime }}$ and $H_{I}=-H_{I}^{^{\prime }}$, the
evolution operator (in the interaction picture) becomes
\begin{equation}
U_{I}^{^{\prime }}\left( \frac{\tau }{2}\right) =U_{I}\left[ \frac{\tau }{2}%
;-g_{k}\right] .
\end{equation}

The combined dynamics from the above two steps is now described by the
following evolution operator
\begin{eqnarray}
\widetilde{U}_{I}\left( \tau \right)  &=&U_{I}^{^{\prime }}\left( \frac{\tau
}{2}\right) U_{I}\left( \frac{\tau }{2}\right)   \nonumber \\
&=&U_{I}\left( \frac{\tau }{2};-g_{k}\right) U_{I}\left( \frac{\tau }{2}%
;g_{k}\right)   \nonumber \\
&=&\exp [-iM(\tau )\sigma _{x1}\sigma _{x2}]\exp [-iN\left( \tau \right) ],
\end{eqnarray}%
where%
\begin{eqnarray}
M(\tau ) &=&A^{^{\prime }}(\frac{\tau }{2})+A(\frac{\tau }{2})  \nonumber \\
&=&\frac{2g_{1}g_{2}}{\Omega }\left( \frac{2}{\Omega }\sin \frac{\Omega \tau
}{2}-\tau \right) ,
\end{eqnarray}%
and%
\begin{eqnarray}
N(\tau ) &=&D^{^{\prime }}(\frac{\tau }{2})+D(\frac{\tau }{2})  \nonumber \\
&=&\frac{g_{1}^{2}+g_{2}^{2}}{\Omega }\left( \frac{2}{\Omega }\sin \frac{%
\Omega \tau }{2}-\tau \right) .
\end{eqnarray}%
Again we see that after the above two successive operations, we realize an
effective controlled rotation of the two qubits, i.e. $U_{xx}\left( \xi
\right) \equiv \exp [-i\xi \sigma _{x1}\sigma _{x2}]$ can be realized by
fixing $\Theta _{k}={\pi }/{2}$ or $\Phi _{xk}={\phi _{0}}/{2}$ and for a
time $t$ satisfying $M(\tau )=\xi $. As a bonus we find that $f_{k}=0$ in
this case, i.e. the single-bit rotation vanishes automatically during the
two-bit rotation. With both of the above approaches, single bit operations
can be cleanly implemented by setting $\Theta _{k}=\pi $ or $g_{k}=0$.

We now consider the implementation of our protocol in realistic experiments.
We note the typical time for gate operation is
\begin{equation}
\tau _{o}=\frac{2\pi \Omega }{g^{2}}=\rho \frac{2\pi }{g},
\end{equation}%
with $g$ to be roughly understood as the order of magnitude of $g_{k}$ and ${%
2\pi }/{g}$ approximately the single qubit operation time. One can increase
the ratio $\rho ={g}/{\Omega }$ to shorten these operation times similar to
other JJ coupling schemes \cite{maklin1,averin1,Han2,kim,zagoskin,you}. It
is important to emphasize, however, most of these other schemes are based on
an adiabatic evolution of the data-bus, thus are limited to a weak coupling,
or, a small $\rho $. In contrast, our protocol is not confined to the
adiabatic dynamics, thus can operate with a larger $\rho $ for a faster
gate. In fact, even when the capacitance of the large junction is much
larger than the other capacitances, the ratio $\frac{g}{\Omega }$ can still
be made large with realistic experimental parameters. For example, if we
take $E_{J}=800$ ($\mu $eV), $E_{c}=10$ ($\mu $eV), $E_{ci}=200$ ($\mu $eV),
$E_{Jk}=200$ ($\mu $eV), $\Phi _{xk}={\phi _{0}}/{2}$ \cite{vion1,mooij}, we
find ${g}/{\Omega }\simeq 0.23$, a limit may be prohibited for other schemes
yet works well for our protocol.

A key requirement for our protocol is the precise control of the operation
time, especially in the second approach which needs a sudden switch of the
magnetic field during each two-qubit operation. However, the second scheme
need not to change the magnitude of the magnetic field but just to switch
the direction of it. Before concluding, we wish to remark that, although our
circuit resembles that of You {\it et. al.} \cite{you}, our operating scheme
and the underlying physics is different. The essential difference is that
our gate does not require an adiabatic operation, thus is faster.

In conclusion, we have presented an efficient protocol for implementing
controlled interactions between two charge-phase qubits. We have discussed
two alternative approaches to realize our protocol. Our scheme seems to work
over a wide parameter range and is faster than most existing protocols. From
the experimental point of view, our protocol seems advantageous as it only
needs the control of one system parameter, that is the external applied
magnetic field. We have adopted a setup involving charge-phase qubit, thus
our protocol is less sensitive to charge fluctuations along with phase
fluctuation \cite{vion1}. In addition, our circuit arrangement can be easily
scaled up to larger number of JJ qubits with similar controls.

{\it We acknowledge the support of the CNSF (grant No.90203018) and the
knowledged Innovation Program (KIP) of the Chinese Academy of Sciences and
the National Fundamental Research Program of China with No 001GB309310.We
also sincerely thank S. Han and L. You for the helpful discussions with him.}

\end{document}